\documentclass[%
 reprint,
 amsmath,amssymb,
 aps,
]{revtex4-1}
\usepackage{lineno}
\usepackage{wrapfig}
\usepackage{multirow}
\usepackage{amsmath}
\usepackage{amsfonts}
\usepackage{amssymb}
\usepackage{ifthen}
\usepackage{natbib, epstopdf}
\usepackage{graphicx}
\usepackage{dcolumn}
\usepackage{bm}
\usepackage[colorlinks=true, pdfstartview=FitV, linkcolor=blue, citecolor=blue, urlcolor=blue,breaklinks=true]{hyperref}
\usepackage{verbatim} 
\renewcommand\Re{\operatorname{\mathbf{Re}}}

\newcommand{\bfig}{\begin{scheme}}
\newcommand{\efig}{\end{scheme}}
\newcommand{\be}{\begin{equation}}
\newcommand{\ee}{\end{equation}}
\newcommand{\bd}{\begin{displaymath}}
\newcommand{\ed}{\end{displaymath}}
\newcommand{\ban}{\begin{eqnarray*}}
\newcommand{\ean}{\end{eqnarray*}}
\newcommand{\ba}{\begin{eqnarray}}
\newcommand{\ea}{\end{eqnarray}}
\newcommand{\dsp}{\displaystyle}
\newcommand{\nno}{\nonumber}
\newcommand{\nnob}{\nonumber \\}
\newcommand{\pd}[2]{\frac{\partial #1}{\partial #2}}
\newcommand{\pdd}[2]{\frac{{\partial}^2 #1}{\partial {#2}^2}}
\newcommand{\pddm}[3]{\frac{{\partial}^2 #1}{\partial {#2} \partial {#3}}}

\newcommand{\dd}{\mathrm{d}}
\newcommand{\ep}{\epsilon^{\prime}}
\newcommand{\ept}{\epsilon^{e\prime}}

\newcommand{\eo}{\epsilon^{o}}
\newcommand{\Pp}{P^{\prime}}
\newcommand{\sgp}{{\sigma}^{\prime}}
\newcommand{\sgo}{{\sigma}^{o}}
\newcommand{\rcp}{\phi_{rcp}}

\begin{document}
\preprint{APS/123-QED}

\title{Asymptotic analysis of stresses near a crack tip in a 2D colloidal packing saturated with liquid}

\author{Arijit Sarkar}
 \altaffiliation{arijit@che.iitb.ac.in}
\author{Mahesh S Tirumkudulu}%
 \email{mahesh@che.iitb.ac.in}
\affiliation{%
Department of Chemical Engineering,\\
Indian Institute of Technology Bombay,\\
Powai, Mumbai 400076, India.
}%
\date{\today}

\begin{abstract}
The consolidation of colloidal particles in drying colloidal dispersions is influenced by various factors such as particle size and shape, and inter-particle potential. The capillary pressure induced by the menisci, formed between the top layer of particles in the packed bed, compresses the bed of particles while the constraints imposed by the boundaries result in tensile stresses in the packing. Presence of flaws or defects in the bed determines its ultimate strength under such circumstances. In this study, we determine the asymptotic stress distribution around a flaw in a two dimensional colloidal packing saturated with liquid and compare the results with those obtained from the full numerical solution of the problem. Using the Griffith's criterion for equilibrium cracks, we relate the critical capillary pressure at equilibrium to the crack size and the mechanical properties of the packed bed.  The analysis also gives the maximum allowable flaw size for obtaining a crack free packing.
\begin{description}
\item[PACS numbers] 82.70.Dd
\end{description}
\end{abstract}

\maketitle

\section{Introduction}
Dried colloidal particle films find use in a number of applications such as tapes for photography and magnetic storage \citep{lew02}, porous coated printer papers, coating vitamin tablets, synthetic opals, photonic crystals \citep{zen02} etc. The macroscopic properties of the film such as its thickness, particle packing and the mechanical strength are influenced by the drying rate, interparticle potential, particle size and shape and the modulus of the particles.
When a dispersion of colloidal particles is dried, the particles concentrate, eventually reaching a close packed concentration. The liquid menisci on the top layer of particles compresses the packing while the substrate resists transverse deformation of the packing. Consequently, transverse tensile stresses  develop in the packing and when these stresses exceed a critical value, the packing cracks resulting in a variety of crack patterns. Such cracks occur not only in thin films such as paints and coatings but also in thick systems and over geophysical length scales such as in the case of dried river beds.

Most of the experimental investigations of the cracking phenomenon in drying colloidal dispersions have focused on the thin film geometry where stresses have been measured using the classical cantilever bending technique \citep{lew02, mahesh05, jhmn1999}. These measurements show that thin films of monodisperse colloidal dispersions containing identical particles crack at a critical stress that is independent of the particle size but varies inversely with the film thickness \citep{chiu93a, chiu93b, mahesh05, mahesh09a}. In almost all cases, the film nucleates multiple cracks with crack spacing that varies linearly with film thickness \citep{mahesh05, morris00}. Experiments also suggest that irrespective of particle size or moduli, each dispersion has a maximum crack free thickness below which the films do not crack. The critical cracking thickness is found to increase with particle size and moduli in the case of hard polymer and metal oxide particles. A number of investigations have also focused on cracking in confined geometries such as capillary tubes where the dispersion dries from one end resulting in a compaction front of packed particles. While the studies in this geometry have mainly focused on crack tip velocity and its relation to the speed of the compaction front \citep{limat95, morris09, Duf03}, it is only recently that Dufresne and co-workers\citep{Duf10} have been successful in imaging the stress variation near the tip of a propagating (interface) crack at the interface of an elastomer and saturated colloidal bed and extract the stress intensity factor from it. The stress decays as inverse square root of the distance from the crack-tip and is in line with the predictions of classical linear theory for fracture in brittle materials.  

On the theoretical front, Routh and Russel \cite{routh99} have derived a constitutive relation relating the macroscopic stress to macroscopic strain in a drying film. They considered the viscoelastic deformation of a pair of identical particles due to contact and interfacial forces and related the strain at the particle level to these forces. Next, they volume averaged the forces over all orientations to arrive at the macroscopic stress versus strain relationship for a drying film. 
In the absence of particle-solvent interfacial tension, the expression for the macroscopic stress tensor \citep{mahesh05} for identical elastic spheres reduces to 
\ba
\sigma_{ij} &=& \delta_{ij} \left\lbrace -P - \frac{GM\phi_{rcp}}{140} \left( \epsilon^2_{mm}+2 \epsilon_{nm} \epsilon_{mn}  \right) \right\rbrace \nnob
&& -\frac{GM\phi_{rcp}} {35} \left( \epsilon_{mm} \epsilon_{ij}+2 \epsilon_{im}\epsilon_{mj}  \right)
\label{eq:41}
\ea
where, $ \epsilon_{ij} $ is the macroscopic strain, $ P $ is the capillary pressure, $\phi_{rcp}$ is the random close packing concentration, $ G $ is the shear modulus of the particles and $ M $ is the number of contacting neighbors. 
The constitutive equation is an improvement over the traditional poroelasticity models \citep{biot1941, biot1955, biot1956} as the former accounts for the nonlinear deformation at the particle level and the influence of particle size, modulus and packing characteristics on the macroscopic deformation field. The model has been successful in predicting not only the stress profile in drying films of both film forming and cracking systems \citep{mahesh04}, but also in predicting many aspects of the cracking mechanism in the latter \citep{mahesh05}. More recently, Russel et. al. \cite{russel08a} have improved on the above relation by adopting the Hertzian contact mechanics at the particle pair level. The final constitutive relation is also non-linear with the stress varying as three halves power of the strain. Using this relation, they determine the capillary pressure necessary either to open an infinite crack in a flawless film or to extend pre-existing flaws of finite lengths. Their results suggest that flaws which are a fraction of the film thickness are sufficient to initiate cracks that would propagate across the sample at pressures modestly greater than obtained from the energy argument. In a related study Man and Russel \cite{russel08b} demonstrate experimentally the role of flaws in nucleating a crack and show that the critical stress obtained from the energy argument only gives the lower bound.
\begin{figure}[ht]
\begin{center}
\includegraphics[scale=0.6]{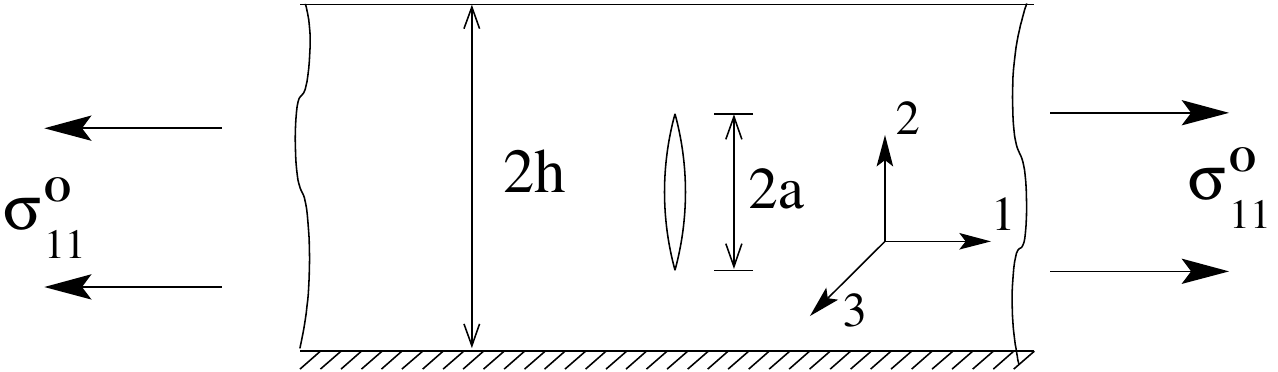} 
\caption {A crack of length $2a$ embedded in a packed bed. The bed is stressed in `1' direction. We consider the case where $a\ll h$.}
\label{crack}
\end{center}
\end{figure}

In this study, we determine the stress field near a crack tip along with the shape of the crack that is present in a two dimensional particle packing saturated with solvent. The flaw is embedded inside the colloidal packing and the size of the flaw is much smaller that any other dimension of packing, say film thickness (Figure \ref{crack}). Further, when the crack dimension in the out of plane direction is larger than $a$, then the only length dimension relevant to the problem is $a$, and the situation becomes amenable to plane stress/strain analysis \cite{suo1992}.
%
%
A packed bed made of an array of colloidal particles can be considered to be a collection of polycrystalline aggregates with pre-existing flaws. These flaws may be attributed to micro-cracks, grain boundaries between the clusters of ordered packing of mono-dispersed particles, dissimilar pores inside the colloidal bed etc. Nucleation of a crack under these circumstances changes the stress field close to the crack with stress concentration at the crack tip. In this work, the stress and strain fields are linearized about the pre-crack state to determine the  disturbance displacement field immediately after the opening of a mode-I crack.  These results also yield the stress intensity factor for the two dimensional elastic field which is then related to the surface energy using the well known Griffith's criterion for equilibrium cracks. The calculated quantities are then compared with the numerical solution for the full problem. The calculations show that the dimensionless critical capillary pressure required to open a crack varies inversely with the crack length to the two thirds' power and depends on a dimensionless parameter that measures the ratio of the elastic to surface energy. 
A simple scaling analysis reveals the essence of the results to follow. Since $\sigma \sim E {\eo}^2$, where `$E$' is effective modulus of the packing and $\eo$ is the characteristic strain in the packing, the elastic energy recovered on the opening of a crack of length `$a$' in a packing of unit thickness is, $\sigma \eo a^2$. Equating this to surface energy ($\gamma a$) and noting that the capillary pressure is linearly related to the stress, gives the critical capillary pressure for opening the crack, $ \frac{P_c R}{2\gamma} \sim A \left( \frac{R}{a} \right)^{2/3} \left( \frac{ER}{\gamma} \right)^{1/3} $, where `$\gamma$' is the surface tension of the solvent, $R$ is the radius of the particles, and $A$ is a constant. The objective of this paper is to rigorously determine the value of $A$ and investigate the consequence of this result.
\section{Nucleation of crack}
\begin{figure}[t]
\begin{center}
\includegraphics[scale=0.38]{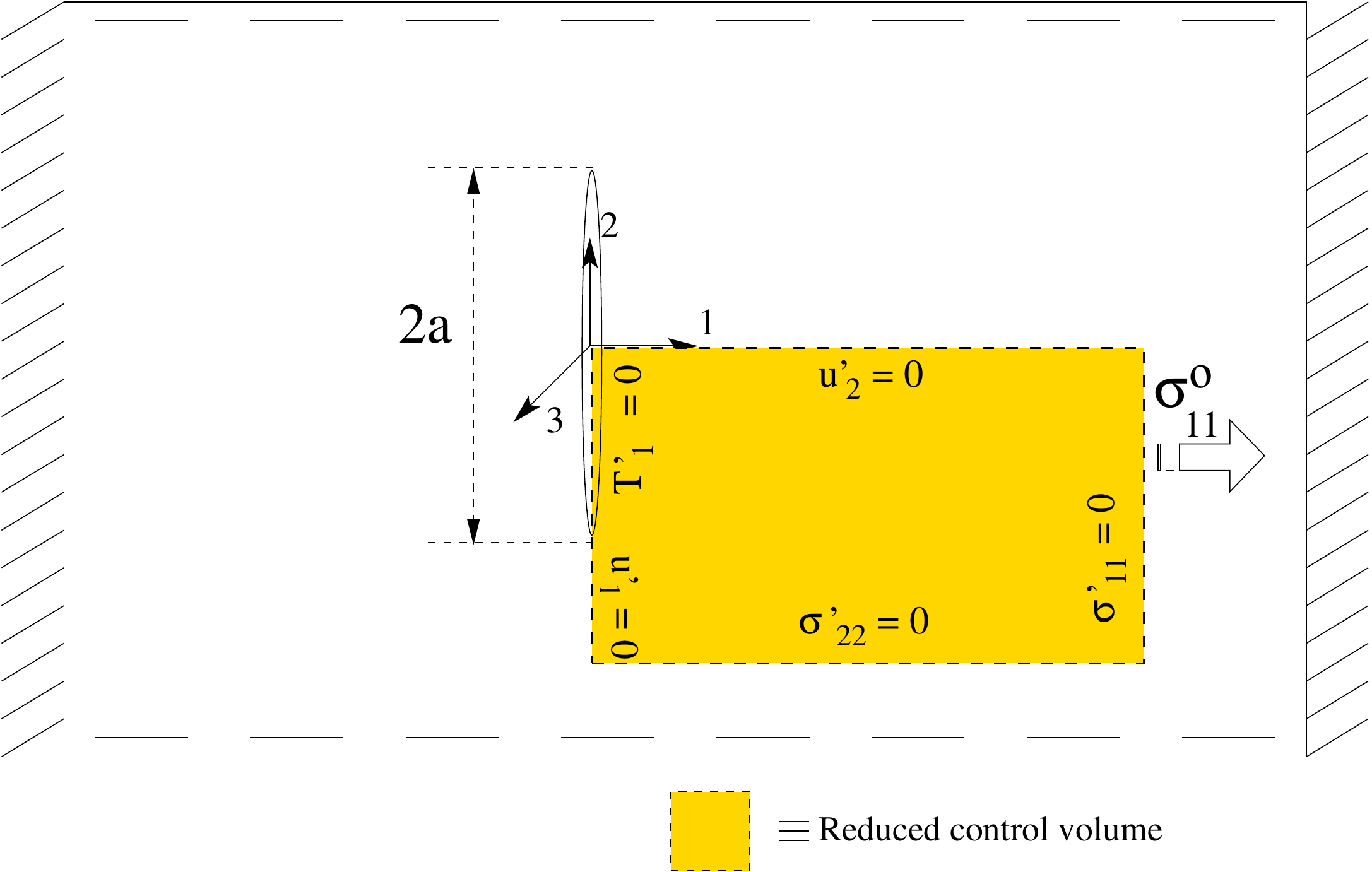}
\caption {A crack of length $2a$ embedded in a packed bed. The bed is stressed in `1' direction. Shaded region shows the region over which the analysis is performed. $T^{\prime}_{1}$ is the traction along crack surface.}
\label{coord}
\end{center}
\end{figure}%
Consider a colloidal packing saturated with water that is confined by solid boundaries at $x_1=\pm L$ with free surfaces at $x_2 = x_3 = \pm h$. As water evaporates, the capillary pressure puts the packing in tension in the $ x_1$ direction so that it is free to contract along $x_2$ and $x_3$ (Figure \ref{coord}). In this case, the strain is given by, $\epsilon_{ij} = - \eo \left( \delta_{i2} \delta_{2j} + \delta_{i3}\delta_{3j} \right) $, where $^o$ denotes the pre-crack values. Volume conservation over a unit volume of the bed relates the strain to the particle volume fraction,  
\begin{equation}
\phi^o = \phi \left( 1-\eo \right)^2 \cdot
\label{eq:42}
\end{equation}
where, $\phi^o$ is the volume fraction in the pre-crack state. The strain and stress fields are sought for a crack with extent $-a < x_2 < a$ in the plain stress formulation,
\ba
&& \epsilon_{ij} = - \eo \left( \delta_{i2} \delta_{2j} + \delta_{i3}\delta_{3j} \right) + \ep_{ij} \;\;\mathrm{ , } \nnob
&& P = P^o + \Pp \;\;\mathrm{ and, }\;\; \sigma_{ij} = \sgo_{ij} + \sgp_{ij} \;\;\mathrm{ , }
\label{eq:43}
\ea
with the perturbed variables represented by the primed quantities and $a<<L,h$. Substituting these in the constitutive equation (\ref{eq:41}) and retaining terms linear in the perturbed quantities gives,
\begin{eqnarray}
&& \bar\sgp_{11} = -\bar \Pp + ( 3\ep_{11} + 2\ep_{22} + 2\ep_{33} ) \nnob
&& \bar\sgp_{22} = -\bar \Pp + ( 2\ep_{11} + 9\ep_{22} + 3\ep_{33} ) \nnob
&& \bar\sgp_{33} = -\bar \Pp + ( 2\ep_{11} + 3\ep_{22} + 9\ep_{33} ) \nnob
&& \bar\sgp_{12} = 4\ep_{12} \nnob
&& \bar\sgp_{23} = 6\ep_{23} \nnob
&& \bar\sgp_{31} = 4\ep_{31} 
\label{eq:4501}
\end{eqnarray}
where a bar over a variable implies a dimensionless quantity with stress and pressure rendered dimensionless with $E \equiv \frac{GM\phi_{rcp} \eo}{35} \cdot$ The dimensionless stress for the pre-crack state is,
\begin{eqnarray}
\bar\sgo_{ij} = \delta_{ij} \left[ - \bar P^o - 2(\eo) \right] - 4(\eo) \left( \delta_{i2} \delta_{2j} + \delta_{i3}\delta_{3j} \right) \cdot
\label{eq:46}
\end{eqnarray}
Since we consider the plane stress case ($\bar \sigma_{3j}=0$) and the bed is stressed only in the $x_1$ direction, $\bar P^o = -6 \eo$, $\bar\sgo_{11}=4 \eo$ and $\bar\sgo_{22}=0 \cdot$

The total amount of particle phase remains constant in the packing, and so the particle volume fractions before and after cracking are related,
\be
\frac{\phi_{rcp}}{\phi} = (1+\ep_{11})(1-\eo+\ep_{22})(1-\eo+\ep_{33}) \quad\mathrm{,}
\label{eq:047}
\ee
where, $\phi_{rcp} $ is the random close packing and the strain is taken to be zero when $\phi=\phi_{rcp}$. The time evolution of stress and strain around the crack can be further subdivided into two limiting cases~\cite{mahesh05} i.e. the short time and the long time limits. In the short time limit, the impact of crack formation on the stress and strain variation is such that it would occur in the absence of solvent flow, suggesting that the material will be incompressible. Thus, in the short time limit and for $ \eo \ll 1$,  (\ref{eq:047}) reduces to,
\begin{equation}
\ep_{11} + \ep_{22} + \ep_{33} = 0.
\label{eq:048}
\end{equation}
Since we shall consider only the plane stress case here ($\bar\sgp_{33} = 0$), $\bar \Pp = \ep_{22} + 7\ep_{33}$. At longer time scales, liquid flows so as to eliminate pressure variations, giving us the required condition for the long time limit, $\bar \Pp =0 $. The perturbed stress and strain are compactly related in the two cases, $\bar\sgp_{ij} = C_{ij} \ept_{ij}$ where $\ept_{ij}$ is the engineering strain and $C$ is the stiffness matrix. The latter is given by,
\be
\quad C = 
\begin{bmatrix} 8	&	6 	&	0\\ 
      	        6	&	12	&	0\\ 
		0	&	0 	& 2 
\end{bmatrix}
\label{eq:049}
\ee
in the short-time limit and,
\be
\quad C = 
\begin{bmatrix} \frac{23}{9} & \frac{4}{3} & 0		 \\ 
                 \frac{4}{3} &     8       & 0     	 \\
                  0          &     0       & 2    
\end{bmatrix}
\label{eq:049a}
\ee
in the long-time limit.
While the original constitutive relation in the pre-crack state (\ref{eq:41}) is for an isotropic solid, (\ref{eq:049}) and (\ref{eq:049a}) imply that the relaxation resulting from the presence of the flaw under the imposed conditions is that for an anisotropic solid. 
A similar observation is noted by Russel and co-workers for the more accurate constitutive relation based on the Hertzian contact mechanics. For convenience, we write the constitutive equations as, $\Delta_{i} = S_{ij} \Sigma_{j}$ where both $\Delta$ and $\Sigma$ are 6-by-1 column vectors. The components of $\Delta$ and ${\Sigma}$ are given by, $\ept_{11}, \ept_{22}, \ept_{33}, \ept_{23}, \ept_{31}, \ept_{12} $ and $\bar \sgp_{11}, \bar \sgp_{22}, ... , \bar \sgp_{12}$ respectively, and the 6-by-6 coefficient matrix, $\mathbf{S}$, is the compliance matrix. The elements of $\mathbf{S}$ are determined from $\mathbf{C}$. 
Note that the present case relates to the case of a orthotropic anisotropy with plane stress condition. Therefore, $\mathbf{S}$ has only seven non-zero elements. 
The displacement field in case of plane strain is easily obtained using the procedure outlined in the next section except that the components of the compliance matrix for the plane stress problem ($S_{ij}$) are replaced with,
\[
D_{ij} = S_{ij} - \frac{S_{i3} S_{j3}  }{S_{33}},\;\;(i,j=1,2,..., 6)
\]
for the plane strain case.
\section{Asymptotic analysis near a crack tip}

The knowledge of the stress fields in the neighborhood of the crack tip is essential in determining the strength of the packed bed. Since the perturbed stress is linear in perturbed strain, we draw upon the mathematical techniques developed in the solid mechanics literature to determine the stress field near the tip of a crack \cite{sih65,lekh81,rice68,esh57,grif21}. The coordinate system ($\tilde x_1,\tilde x_2$) for this analysis is shown in Figure \ref{coord} where the origin is placed at the crack tip so that $\tilde x_1=\bar x_1, \tilde x_2 = \bar x_2 + \bar a$ and the variables have been rendered dimensionless using the characteristic length of the solution domain. The momentum balance equation in $\tilde x_1$ and $\tilde x_2$ directions in the absence of body forces are given by, 
\begin{align}
& \pd{\bar\sgp_{11}}{\tilde x_1} + \pd{\bar\sgp_{12}}{\tilde x_2} = 0,\;\mathrm{and} \nnob
& \pd{\bar\sgp_{21}}{\tilde x_1} + \pd{\bar\sgp_{22}}{\tilde x_2} = 0. 
\label{eq:49}
\end{align}

Following Sih et al.\cite{sih65} and Hoenig\cite{hoenig82},  we related the stresses to stress correlation functions, $\chi$ and $\psi$, via
\begin{align}
\bar\sgp_{ij} &= - \pddm{\chi}{\tilde x_i}{\tilde x_j} + \delta_{ij} \pdd{\chi}{\tilde x_m} ,  \;\mathrm{and}\nnob
\bar\sgp_{3i} &= e_{ij} \pd{\psi}{\tilde x_j} 
\label{eq:410}
\end{align}
with $e_{ij}$ and $\delta_{ij}$ being the second order alternating and Dirac delta tensor respectively, and $i,j$ allowed values of 1 or 2. Note that (\ref{eq:410}) automatically satisfies (\ref{eq:49}). The above expressions along with the constitutive relation are substituted in the compatibility equations,
\ba
& & \pdd{\ept_{11}}{\tilde x_2} + \pdd{\ept_{22}}{\tilde x_1} - \pddm{\ept_{12}}{\tilde x_1}{\tilde x_2} =  0,\; \mathrm{and} \nnob
& & \pddm{\ept_{11}}{\tilde x_2}{\tilde x_3} = \pddm{\ept_{12}}{\tilde x_3}{\tilde x_1} - \pdd{\ept_{23}}{\tilde x_1} + \pddm{\ept_{31}}{\tilde x_1}{\tilde x_2} 
\ea
to give, respectively,
\begin{align}
& S_{11}\chi_{,2222} + (2S_{12} + S_{66})\chi_{,1122} + S_{22}\chi_{,1111} = 0,  
\quad \mathrm{and} \label{eq:4101} \\
& S_{44} \psi_{,11} + S_{55}\psi_{,22} = 0.
\label{eq:4102}
\end{align}
The degree of anisotropy in the material can be judged by rewriting (\ref{eq:4101}) differently,
\be
\nabla^4{\chi} + \delta_1 \chi_{,1111} +  \delta_2 \chi_{,2222} = 0 
\label{eq:4103}
\ee
where $1 + \delta_{1}= \frac{2S_{22}}{S_{66} + 2S_{12}}$ and $ 1 + \delta_2 = \frac{2S_{11}}{S_{66} + 2S_{12}}$  are indicators of anisotropy in the material. The difference in the values of $\delta_1$ and $\delta_2$ originate from the fact that the bed is held along `1' direction and perturbed along `2' and `3' directions leading to a directional perturbation of the stress field. When $|\delta_{i}| \ll 1$, $\chi$ satisfies the biharmonic equation, i.e. the material is isotropic. In the current problem, $\delta_i$ are $ -0.12 $ and $ 0.33 $ in the short-time limit and $ -0.24 $ and $1.4 $ in the long-time limit suggesting that  the anisotropy is significant and cannot be ignored.

(\ref{eq:4101}) and (\ref{eq:4102}) are a pair of decoupled equations in $\chi$ and $\psi$,
\begin{eqnarray}
L_{4} \chi &=& 0 \\ 
L_{2}\psi &=& 0 
\label{eq:411}
\end{eqnarray}
where  the differential operators are given by, $L_2 \equiv S_{44} \pdd{}{ \tilde x_1} + S_{55} \pdd{}{ \tilde x_2}$ and $L_4 \equiv S_{11} \frac{\partial^4}{\partial  \tilde x_2^4} + 2(S_{12}+S_{66}) \frac{\partial^4}{\partial \tilde x_1^2 \partial  \tilde  x_2^2} + S_{22} \frac{\partial^4}{\partial  \tilde x_1^4}$.
Lekhnitskii\cite{lekh81} has shown that $L_2$ and $L_4$ can be decomposed into two and four linear operators of first order respectively, of the form $D_k= \partial/ \partial  \tilde x_2 - \mu_k \partial / \partial  \tilde x_1$ such that $D_1 D_2 D_3 D_4 \chi =0$ and $D_5 D_6 \psi = 0$. Substitution of $D_k$ in $L_4$ and $L_2$ shows that $\mu_k$ are roots of the  polynomial operators, $l_4 \equiv S_{11}\mu_k^4 +  (2S_{12}+S_{66})\mu_k^2 + S_{22}=0$ and $l_2 \equiv S_{55}\mu_k^2 + S_{44}=0$.
Then, the stress correlation functions can be written as,
\ba
\chi & = & \dsp\sum^{2}_{i=1} \left \{ \chi_i( \tilde x_1 + \mu_i  \tilde x_2) + \chi_i( \tilde x_1 + \bar \mu_i  \tilde x_2) \right \} \quad \mathrm{and} \nnob
\psi & = & \psi ( \tilde x_1 + \mu_i  \tilde x_2) + \psi ( \tilde x_1 + \bar \mu_i  \tilde x_2)
\ea
where the bar on $\mu_i$ represents the conjugate complex number. Further, Lekhnitskii\cite{lekh81} has shown that for the elastic energy of the packing to be positive, the roots cannot be real.
Consequently, the general expression for the stress function will involve the real part of both the complex conjugates,
\begin{eqnarray}
\chi_R( \tilde x_1, \tilde x_2) &=& 2 \Re \left\lbrace  \dsp \sum^{2}_{i=1} \chi_i(z_i) \right\rbrace  \nnob
\psi_R( \tilde x_1, \tilde x_2) &=& 2 \Re \left\lbrace  \psi(z_3) \right\rbrace 
\label{eq:412}
\end{eqnarray}
where $z_i = \tilde x_1 + \bar \mu_i  \tilde x_2$ and $\mathbf{Re}$ represents the real part. Since the stresses are related to the derivatives of the above functions, it is convenient to assume the following functional form,
\ba
&& \frac{ \partial \chi_k (z_k)}{\partial z_k} =  G_k(z_k)\; \mathrm{for}\; k\in[1,2] \nnob
\mathrm{and}\; && \psi_3(z_3) = G_3(z_3)  
\ea
so that the stresses are given by,
\begin{align}
& \bar\sgp_{11} = \pdd{\chi_R}{\tilde x_2} = 2\Re\left [\mu^2_1 \frac{\mathrm{d}G_1}{\dd z_1} + \mu^2_2 \frac{\mathrm{d}G_2}{\dd z_2} \right ], \nnob
& \bar\sgp_{22} = \pdd{\chi_R}{\tilde x_1} = 2\Re \left [\frac{\dd G_1}{\dd z_1} + \frac{\dd G_2}{\dd z_2} \right ], \nnob
& \bar\sgp_{12} = -\pddm{\chi_R}{\tilde x_1}{\tilde  x_2} = -2\Re \left [\mu_1 \frac{\dd G_1}{\dd z_1} + \mu_2 \frac{\dd G_2}{\dd z_2} \right ], \nnob
& \bar\sgp_{31} = \pd{\psi_R}{\tilde x_2} = 2\Re \left [\mu_3 \frac{\dd G_3}{\dd z_3} \right ],  \;\mathrm{and}\nnob
& \bar\sgp_{23} = -\pd{\psi_R}{\tilde x_1} = -2\Re \left  [\frac{\dd G_3}{\dd z_3} \right ]. 
\label{eq:4121}
\end{align}
We can also determine the strain components in terms of the stress function, for example, 
\ba
\ept_{11} &=&  \pd{\bar u^{\prime}_1}{\tilde x_1} \nnob
&=& 2 \Re \left [ \frac{\dd G_1}{\dd z_1} (S_{11}\mu^2_1 + S_{12} - S_{16}\mu_1) \right ] \nnob
  && + 2 \Re \left [ \frac{\dd G_2}{\dd z_2} (S_{11}\mu^2_2 + S_{12} - S_{16}\mu_2) \right ] \cdot
\ea
Integrating the above expression with respect to $z_j$ gives the displacement field in the `1' direction,
\ba
\bar u^{\prime}_1 &=& 2 \Re \left\lbrace  \dsp\sum^{2}_{j=1} p_{1j} G_j(z_j) \right\rbrace
\ea
where, $p_{1j} = S_{11}\mu^2_j + S_{12} - S_{16}\mu_j \cdot$ \\
Following a similar procedure for the remaining strain components, all displacements are determined,
\begin{equation}
\bar u^{\prime}_i = 2 \Re \left\lbrace  \dsp \sum^{3}_{j=1} p_{ij} G_j (z_j) \right\rbrace 
\label{eq:41201}
\end{equation}%
where 
\begin{eqnarray}
p_{1i} &=& S_{11}\mu_i^2 + S_{12} - S_{16}\mu_i \nnob
p_{2i} &=& S_{12}\mu_i + S_{22}/\mu_i - S_{26} \nnob 
p_{33} &=& S_{45} - S_{44}/\mu_3 \nnob
p_{31} &=& p_{32} = p_{13} = p_{23} = 0
\label{eq:41202}
\end{eqnarray} 
Next, we ascertain the functional form of $G_i$. Rice\cite{rice68} has shown that the J-integral, \\
$ J = \dsp\int_{{\dd}{\Omega}} \left( W \dd \tilde x_1 - \bar \sgp_{ij} n_j \frac{\dd \bar u_i}{\dd \tilde x_2} \dd S \right)$, has the same value for all integration paths surrounding crack tips in two dimensional fields of linear or nonlinear elastic materials. Here, $ W $ is the strain energy density, $ n_j $ is the normal to the chosen path, and $S$ is the distance along the path ${\dd}{\Omega}$. Assuming $\frac{\dd G_i}{\dd z_i} \varpropto {z_i}^{p}$ in the neighbourhood of the crack opening, $W \varpropto {z_i}^{2p}$ and  $\bar \sgp_{ij} n_j \frac{\dd \bar u_i}{\dd \tilde x_2} \sim {z_i}^{2p}$.
Hence, $ J \sim {z_i}^{2p+1} $. Since the value of $J$ should be independent of the path, $ p = - \frac{1}{2} $. Thus, we assume $G = B_i \sqrt{{2 \bar a z_i}/{\pi}}$ for a flaw of size $\bar a $ where $B_i$ is the stress function amplitude. 
\begin{figure}[t]
\begin{center}
\includegraphics[scale=0.80]{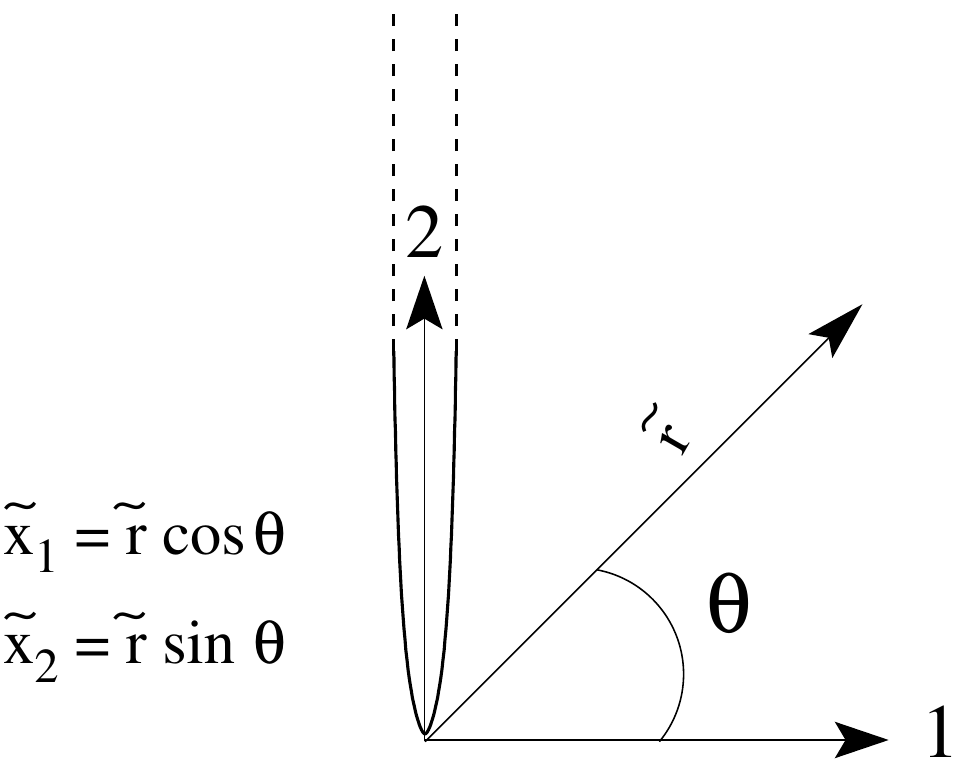}
\caption {Shifted coordinate system for the asymptotic analysis}
\label{coasymp}
\end{center}
\end{figure}
For a \underline{stress free} crack surface, the stress and displacement components near the crack tip become,
\begin{align}
\bar\sgp_{11} &= \sqrt{\frac{2\bar  a}{\pi \tilde r}} \Re \dsp\sum^{2}_{i=1} \frac{B_i \mu_i^2}{\sqrt{\cos\theta + \mu_i \sin\theta}},\nnob
\bar\sgp_{22} &= \sqrt{\frac{2\bar  a}{\pi \tilde  r}} \Re \dsp\sum^{2}_{i=1} \frac{B_i}{\sqrt{\cos\theta + \mu_i \sin\theta}}, \nnob
\bar\sgp_{12} &= -\sqrt{\frac{2\bar  a}{\pi \tilde r}} \Re \dsp\sum^{2}_{i=1} \frac{B_i}{\sqrt{\cos\theta + \mu_i \sin\theta}},\nnob
\bar\sgp_{31} &= \sqrt{\frac{2\bar a}{\pi \tilde r}} \Re \frac{B_3 \mu_3}{\sqrt{\cos\theta + \mu_3 \sin\theta}}, \nnob
\bar\sgp_{23} &= -\sqrt{\frac{2\bar a}{\pi \tilde r}} \Re \frac{B_3}{\sqrt{\cos\theta + \mu_3 \sin\theta}}, \; \mathrm{and} \nnob
\bar u^{\prime}_i &= 2\sqrt{\frac{2\bar a\tilde r}{\pi}} \Re \dsp\sum^{3}_{j=1} p_{ij} B_j{\sqrt{\cos\theta + \mu_j \sin\theta}}.
\label{eq:41301}
\end{align}
Note that the perturbed normal stress at the crack surface for the current problem has a finite non-zero value ($-\bar \sgo_{11}$) which will require minor modifications to some of the above expressions and is dealt with towards the end of this section.
The stress intensity factor ($K$) for mode I crack is defined as,
\be
\bar K_1 = \Re{\bar{K}_1} = \dsp \lim_{\substack{{\tilde x_2}\rightarrow0^- \\ {\tilde x_1} = 0 }} \bar \sgp_{11} \sqrt{2 \pi \tilde r} 
= 2 \sqrt{\bar a} \Re  \dsp\sum^{2}_{i=1} \frac{B_i \mu^2_i}{\sqrt{-\mu_i}}
\label{eq:414}
\ee
Similarly, $\bar K_2$ and $\bar K_3 $ can be obtained from $ \bar \sgp_{12} $ and $ \bar \sgp_{13}$. Thus the stress intensity factor for the three modes can be written compactly, 
\be
\mathbf{\bar{K}} = -2 \mathbf{i} \sqrt{\bar a} \mathbf{N} \mathbf{I_{\mu} B}
\label{eq:41401}
\ee
where $i=\sqrt{-1} \;,$
\be [N] = \begin{bmatrix} \mu^2_1 & \mu^2_2 &  0  \\ 
	                    -\mu_1   & -\mu_2  &  0  \\ 
	                        0    &    0    &\mu_3\\
       	     \end{bmatrix}
 ,\; \mathrm{and} \quad [I_{\mu}] = \begin{bmatrix} \frac{1}{\sqrt{\mu_1}} & 0		&  0	      \\ 
	                   	          0    & \frac{1}{\sqrt{\mu_2}}  &  0	      \\ 
	                       	          0    &    0           &\frac{1}{\sqrt{\mu_3}}\\
       	     \end{bmatrix} \cdot
\label{eq:41402}
\ee
The perturbed displacements of the crack surface can be found in terms of the distance from tip along the crack surface, $\theta \rightarrow \frac{\pi}{2}$, $\tilde r=\tilde \zeta$
\be
\mathbf{\bar u}^{\prime} = \mp \sqrt{\frac{2\tilde \zeta}{\pi}} \mathbf{Q}^{-1} \mathbf{\bar K}
\label{eq:41601}
\ee%
where $\mathbf{Q^{-1}} =\mathbf{Im} \left \{ \mathbf{pN^{-1} I^{-2}_{\mu}} \right \}$.

The above analysis gives the functional form of the stress and strain fields close to the crack tip in terms of the unknown $\bar{K}$. In order to determine the stress intensity factor, we assume a finite sized crack with an elliptical shape such that the minor axis of dimensionless length $2\bar  c$ is small compared to the major axis ($2 \bar a$), $\alpha = \frac{\bar  c}{\bar a} \ll 1$. Eshelby \cite{eshelby57} has shown that for an elliptical crack, the strain is uniform around the crack. Following Hoenig\cite{hoenig82}, the displacement of the crack surface is given by ($i=1,2,3$), 
\begin{eqnarray}
\bar U^{\prime}_i =  A_{1i} \bar{x}_1 = \beta_i \sqrt{\bar a^2- {\bar{x}_2}^2 }\;\; \mathrm{where, } \;\; \beta_i = \frac{A_{1i}}{\alpha},\nno 
\label{eq:416}
\end{eqnarray}
and the strains by,
\[
\ept_{11} = \frac{\beta_1}{\alpha}, \; \ept_{12} =  \frac{\beta_2}{2\alpha},\; \mathrm{and}\;  \ept_{13}  = \frac{\beta_3}{2\alpha}.
\]
Thus, the perturbed stress at the crack face for this simple crack model is related to $\beta_i$, $\bar \sgp_{1k} = C_{k l} \beta_{l} $.
Here,  the origin of the coordinate system ($\bar{x}_1$,$\bar{x}_2$) lies at the center of the ellipse with $\bar x_2$ directed along the major axis (Figure \ref{coord}). Writing the crack face displacement in terms of the coordinate system with the origin placed at the crack tip (Figure \ref{coasymp}),
\be
\bar U^{\prime}_i = \beta_i \sqrt{2\bar a\tilde \zeta}
\label{eq:417}
\ee
where $\tilde \zeta$ is the distance from the crack tip along $\tilde x_2$. 
Comparing (\ref{eq:41601}) with (\ref{eq:417}) we get 
\be
\mathbf{Q}^{-1} \mathbf{\bar K} = [\beta] \sqrt{\pi\bar  a}.
\label{eq:41701}
\ee
which relates the unknowns, $\mathbf{\bar K}$ and $[\beta]$. In order to complete the problem, we determine the elastic  energy released from the simple crack model, 
\be
\bar \xi = 2 \frac{1}{2} \int^{\bar a}_{-\bar a} \bar\sigma^{o}_{1k} \bar U_k \dd{\tilde x_2} = -C_{ik} \beta_i \beta_k \frac{\pi \bar a^2}{2}
\label{eq:4163}
\ee
and equate $\frac{\dd \bar \xi}{\dd \bar a} = 2 \bar J$ giving,
\be
\bar K_i = \sqrt{\pi \bar  a} \bar \sigma^{o}_{1j}
\label{eq:41631}
\ee
where $\bar J$ is value of the standard $J$-integral \citep{rice68}, determined in the limit as the integration path is shrunk so as to lie along the crack face,
\be
\bar J = \dsp\lim_{\delta \rightarrow 0} \frac{1}{\delta} 
\dsp\int^{\delta}_{0} \bar \sgp_{1i}(\delta-\tilde r, -\frac{\pi}{2}) \bar u_{i}^{\prime}(\tilde r,\frac{\pi}{2}) \dd \tilde r  = -\frac{1}{2} \bar K_i \left( Q^{-1}_{il} \bar K_l \right). 
\label{eq:41632}
\ee
Comparing (\ref{eq:41701}) and (\ref{eq:41631}) gives the expression for $\beta_i$ in terms of the far field stresses,
\[\mathbf{\beta}_i = \mathbf{Q}_{ik}^{-1} \mathbf{\bar\sgo}_{1k} \cdot\]
%
%
We are now in a position to write the elastic energy recovered (dimensional) due to the formation of a finite length mode-I crack of length $a$,
\be
\xi = -\frac{\pi}{2} a^2 Q^{-1}_{11} \dfrac{(\sigma^{o}_{11})^2}{E}
\label{eq:41633}
\ee

The present problem requires the crack surface to have a normal stress ( $- \bar\sgo_{11}$ ) and the far field perturbed stress to be zero. Consequently, the  complex stress function $G_{k}(z_k)$ in (\ref{eq:4121}) is replaced with $G_{k} (z_k)+ \Gamma_k z_k$ ($k=1,2$), where $\Gamma_k$ are real constants \cite{sih65}. Substituting the new expression for $G_{k}$ and applying the traction boundary condition at the crack surface, we get,
\be	
\begin{bmatrix} \Gamma_1   \\ 
	                    \Gamma_2      \\ 
	 \end{bmatrix}
	 =
	 \mathbf{Re} \left \{ \frac{1}{\mu_1 \mu_2 (\mu_1 - \mu_2)}
	\begin{bmatrix}  	\mu_2  & \mu_2^2	      \\ 
	                   	          \mu_1 & \mu_1^2  	      \\ 
	\end{bmatrix} 
	\begin{bmatrix} -\bar \sgo_{11}   \\ 
	                    0      \\ 
	 \end{bmatrix}
	 \right\}
\ee
The above expression along with the definition for the stresses (\ref{eq:4121}) suggests that both $\bar\sgp_{11}$ and $\bar \sgp_{22}$ are influenced by the traction condition. However, neither $\bar\sgp_{12}$ nor $\bar u^{\prime}_{1}$ along the crack face are effected implying that the energy calculations and the corresponding values of the stress intensity factors remain unchanged.

The total energy for the system is $\mathcal{E} = \xi + \Gamma$, where $\Gamma=4\gamma a $ corresponds to surface energy and $\gamma$ is the surface tension.
Following Griffith's argument \citep{grif21}, the crack will be in equilibrium when, 
\ba
&& \frac{\dd \mathcal{E}}{\dd a} = 0 \Rightarrow \sigma^{o}_{11} = \sqrt{\dfrac{4\gamma E}{Q^{-1}_{11} a \pi}}
\ea
which relates the far field stress to the crack length and surface tension. Since $E$ is a linear function of the pre-crack strain which in turn is related to the far field stress, we have
\be
\sigma^{o}_{11} (a)^{2/3} = \left[ \frac{2\gamma}{Q^{-1}_{11} \pi} \right]^{2/3} \left( \frac{G M \phi_{rcp}}{35}\right)^{1/3}
\ee
A more useful relation is in terms of the capillary pressure,
\be
\left (-\frac{P^o R}{2 \gamma} \right )\left (\frac{a}{R} \right) ^{2/3} =\left \{ \frac{3}{4} \left[ \frac{ 8}{35 Q^{-2}_{11} \pi^2} \right]^{1/3}\right \} \left( \frac{G M \phi_{rcp}R }{2 \gamma} \right)^{1/3} .
\ee
Thus the dimensionless critical capillary pressure is related to the dimensionless crack length,
\be
 (-\tilde P^o)(\tilde a^{2/3}) = A W^{1/3}
\label{eq:41634}
\ee
where, $W=G M \phi_{rcp} R/ 2 \gamma$ represents the balance of the elastic and surface energy and $A$ is equal to 0.45 and 0.35 for the short and long time limit, respectively.


\section{Numerical solution}
The stress and the displacement fields obtained in the previous section is applicable to regions close to the crack tip. For the full solution, the momentum balance equations (\ref{eq:49}) for the perturbed stresses were solved numerically for the control volume highlighted in Figure \ref{coord} using finite element method (DIFFPACK$\textsuperscript{\textregistered}$). The boundary conditions are as follows,
\ban
 \bar \sgp_{11} = & -\bar  \sigma^o_{11}\;  &\mathrm{for}\; \bar x_1 = 0, 0>\bar x_2> -\bar  a, \\ \nno
\bar u^{\prime}_1  = & 0\; &\mathrm{for}\; \bar x_1=0, -\tilde a >\bar x_2>- 1,  \\ \nno
 \bar u^{\prime}_2  = & 0\; &\mathrm{for}\;0< \bar x_1<1, \bar x_2=0,  \\ \nno
  \bar \sgp_{22} = & 0 \; &\mathrm{for}\;0< \bar x_1<1, \bar x_2=-1,\;\mathrm{and}  \\ \nno 
 \bar \sgp_{11} = & 0\; & \mathrm{for}\; \bar x_1 = 1, 0>\bar x_2> -1 \nno 
\ean
where $\bar a <<1$ so that the stress and the strain fields close to the cracks are not influenced by the size of the control volume. 

Rectangular elements were used with adaptive refinement of the grid near the crack tip. The total number of nodes in the control volume were about 50,000.
\section{Results and Discussion}
\subsection{Numerical Simulation}

\begin{figure}[t]
\begin{center}
\includegraphics[scale=0.28]{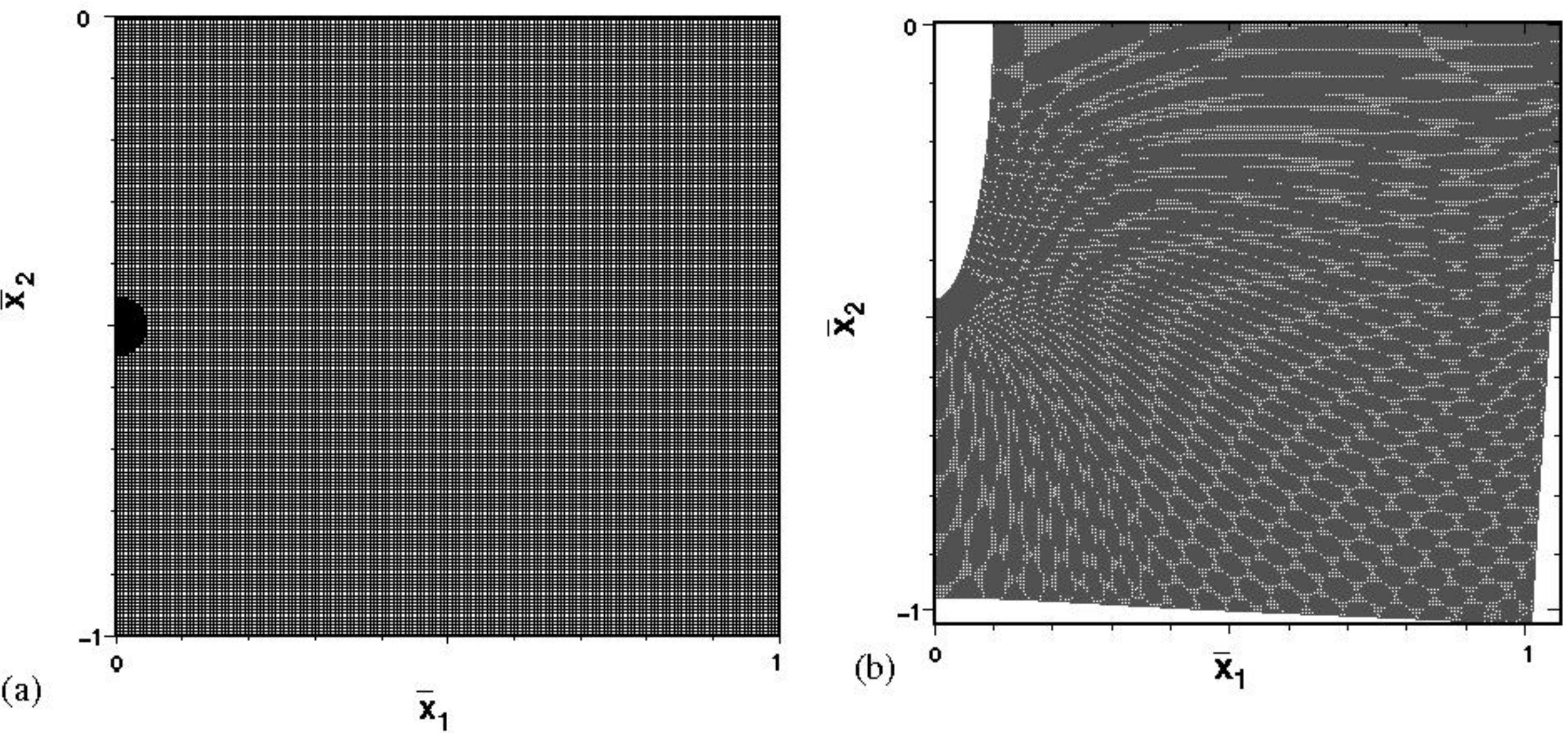} 
\caption {(b) The basic grid before deformation. (b) The scaled surface deformation of the control volume for the short time limit. All displacements have been scaled with  one-tenth the maximum displacement ($u^{\prime}_1(0,0)$).}
\label{grid}
\end{center}
\end{figure}

\begin{figure}[t]
\begin{center}
\includegraphics[scale=0.28]{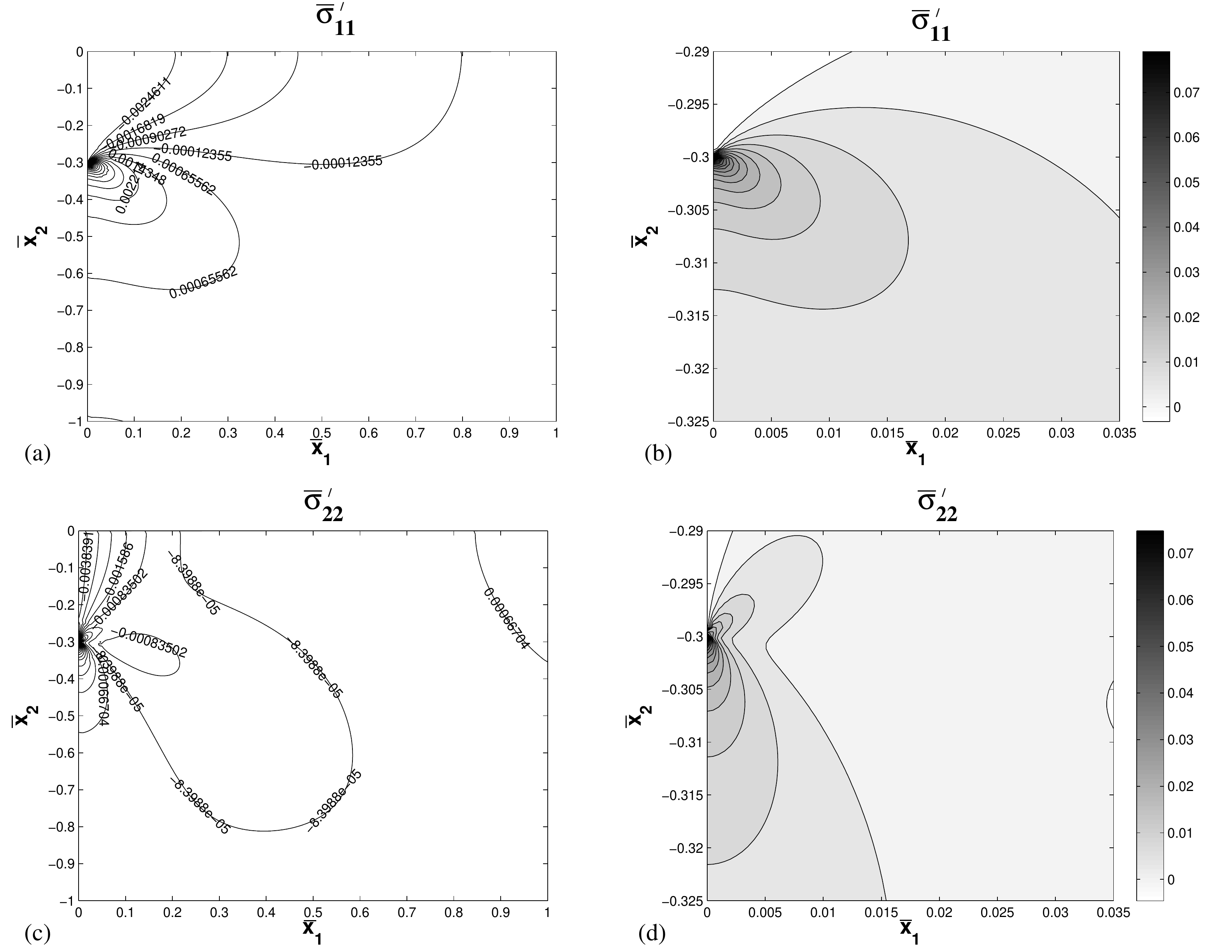} 
\caption {Variation of perturbed stresses for nondimensional half crack length, $\bar a = 0.3$ and $\eo = 7.8\times10^{-4}$ : (a) Contour plot of $\bar\sgp_{11}$, (b) Gray scale plot of $\bar\sgp_{11}$ close to crack tip, (c) contour plot of $\bar\sgp_{22}$,  and (c) Gray scale plot of $\bar\sgp_{11}$ close to crack tip.}
\label{sigma2233}
\end{center}
\end{figure}

\begin{figure}[t]
\begin{center}
\includegraphics[scale=0.28]{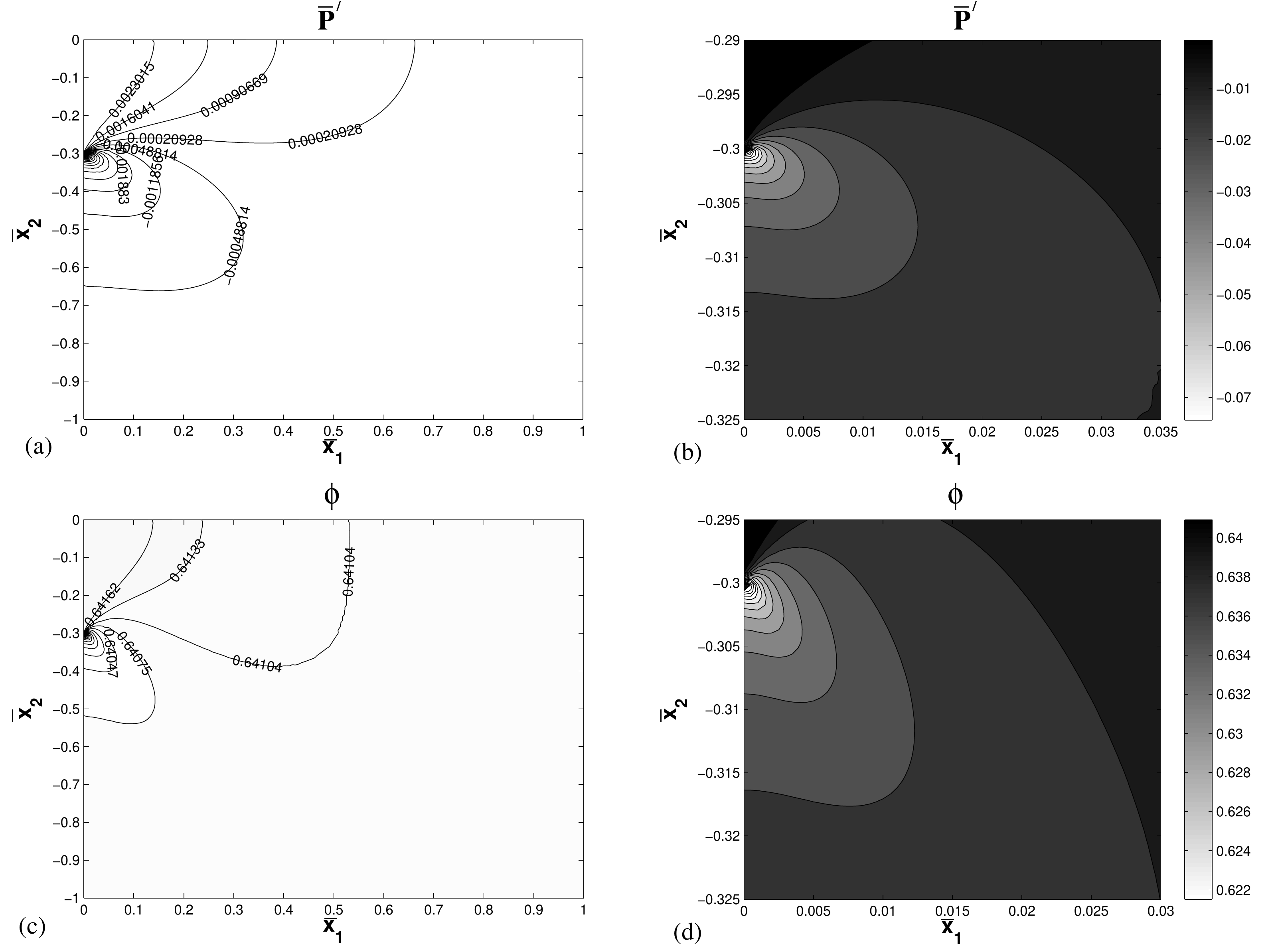} 
\caption {Variation of perturbed pressure in the short time limit for $\bar a = 0.3$ : (a) Contour plot of $\bar \Pp$, (b) Gray scale plot of $\bar \Pp$ close to crack tip. Variation of the particle concentration in the long time limit: (c) Contour plot of $\phi$, and (d) Gray scale plot of $\phi$ close to crack tip.}
\label{pru2}
\end{center}
\end{figure}

The perturbed stress, strain and displacements obtained from the numerical simulations are presented in Figure \ref{grid}-\ref{pru2}. Unless specified, all results pertain to the short time limit. Figure \ref{grid}(a) and (b) presents the initial and deformed grid, respectively. The surface displacements in Figure \ref{grid}(b) have been scaled so as to highlight their magnitude.
As expected, the displacement at the center of the crack is maximum with the tip of the crack moving upwards. Since the perturbed stresses are zero at the control volume boundaries, the effect of the surface displacement at the crack faces can be observed at the boundaries. The equilibrium shape of the crack surface is elliptical and the ratio of the length of the minor to major axis is very small ($\sim10^{-3}$), both of which are in agreement with the asymptotic solution.
Figure \ref{sigma2233} presents the simulated values of $\bar \sgp_{11}$ and $\bar \sgp_{22}$ for the control volume. The contour plot (Figure \ref{sigma2233}(a) and (c)) and the gray scale plots of the region close to the crack tip (Figure \ref{sigma2233}(b) and (d)) demonstrate the sharp decrease in stress with increasing distance from the crack tip. The contours are perpendicular to the symmetry surfaces ($\bar x_2 = 0$ and $-0.3<\bar x_1<1, \bar x_2 = 0$) as expected from the boundary conditions while they decay to zero at $\bar x_1=1$ and $\bar x_2 =-1$. Figure \ref{pru2}(a) and (b) present the perturbed pressure for the short time limit. Interestingly, the pressure is negative close to the tip suggesting that the solvent will flow towards the tip once the crack nucleates. This is borne out in the simulations for the long time limit (Figure \ref{pru2}(c) and (d)) where the particle concentration has reduced at the tip. Note that while the particle concentration should always be equal or more than the close pack concentration at all times, values of $\phi<\phi_{rcp}$ near the crack tip in the long time limit are not physical since no such constraint has been imposed in the simulation. Instead, extra solvent could accumulate at the crack tip between the crack faces.
\subsection{Comparison with Asymptotic Solution}
\begin{figure}[t]
\begin{center}
\includegraphics[scale=0.35]{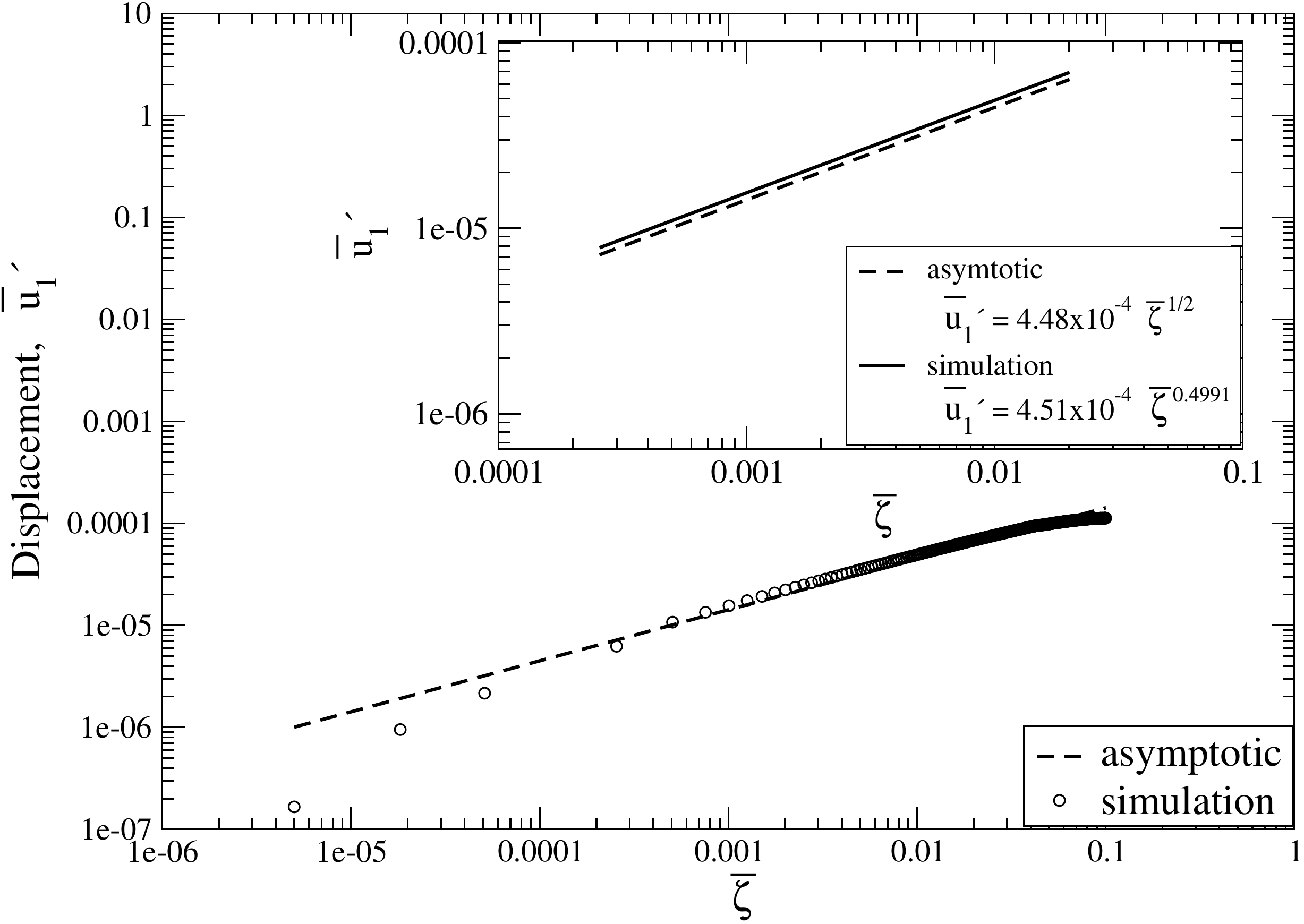}
\caption {Displacement $\bar u^{\prime}_1$ for $\bar a=0.3$ along the crack face.}
\label{short_t_u}
\end{center}
\end{figure}

\begin{figure}[htp]
\begin{center}
\includegraphics[scale=0.35]{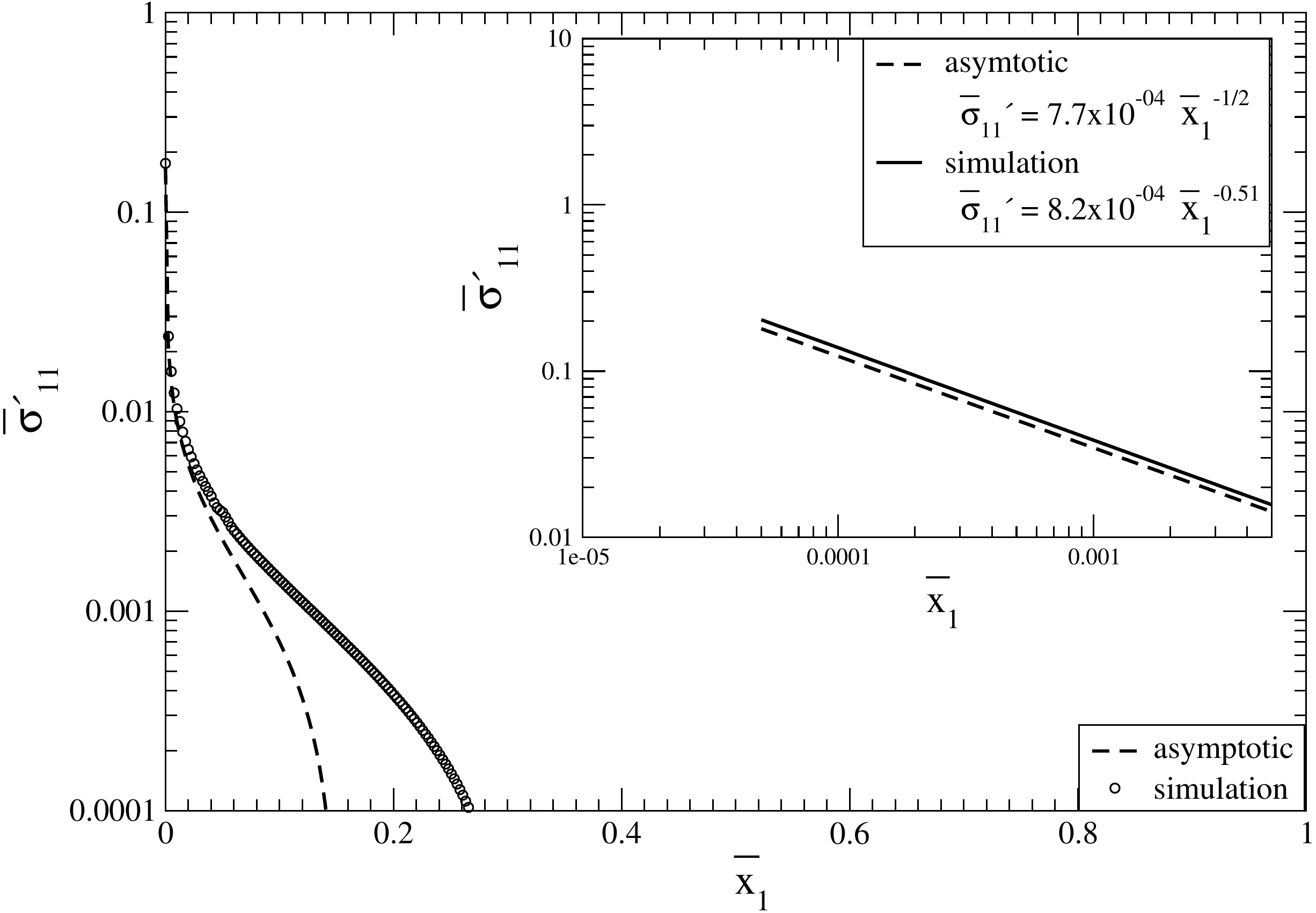} 
\caption {Perturbed stress $\bar \sgp_{11}$ along $\bar{x} _1$ ($\bar x_2 = -\bar a$) for $\bar a=0.3$.}
\label{short_t_sigma22}
\end{center}
\end{figure}

\begin{figure}[htp]
\begin{center}
\includegraphics[scale=0.32]{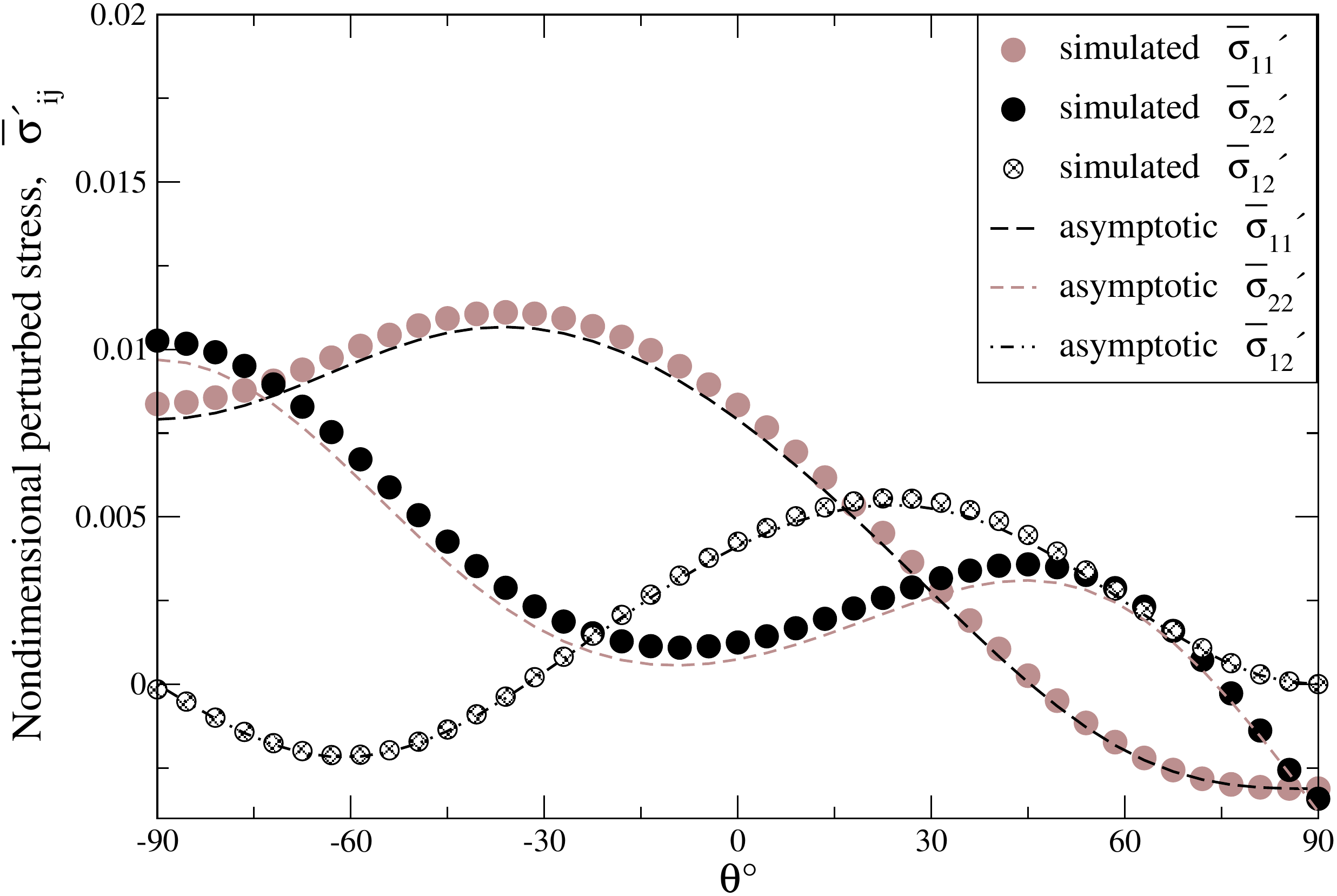}
\end{center}
\caption
{Angular variation of perturbed stresses for short time  at $|\bar r - \bar a| = 0.012$ for $\bar a =0.1$.}
\label{sigma_theta_short}
\end{figure}

\begin{figure}[htp]
\begin{center}
\includegraphics[scale=0.37]{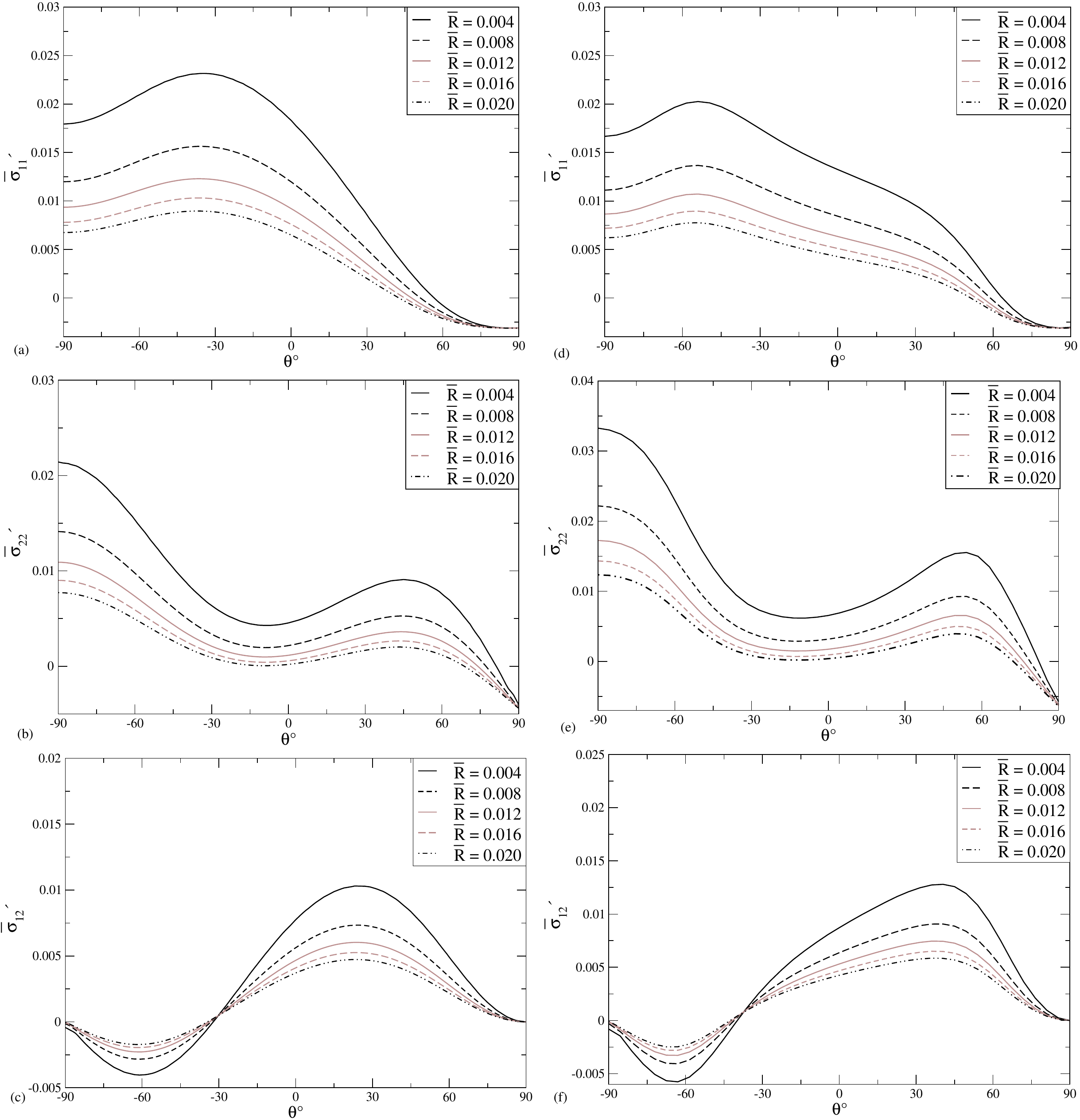}
\caption {Angular variation of perturbed stresses at different radii ($\bar R = \bar r - \bar a $), (a)--(c) in the short time limit, and (d)--(f) in the long time limit for $\bar a=0.3$.}
\label{short_t_sigma}
\end{center}
\end{figure}

Figure \ref{short_t_u} compares the spatial variation of perturbed displacement, $\bar u^{\prime}_1$, along the crack face obtained from the simulation with that predicted by the asymptotic solution. At the crack tip, $\bar u^{\prime}_1=0$ while for $10^{-4}<\bar x_1< 10^{-2}$, the displacement varies as the square root of the distance from the tip. The disagreement close to and far away from the crack tip is attributed to the limitation on grid refinement in case of numerical solution close to the tip (which is unable to capture the large variations in the stress) and to the non-applicability of the asymptotic solution far away from the crack tip.  
Figure \ref{short_t_sigma22} presents the spatial variation of $\bar\sgp_{11} $ along $\bar x_1$ away from the crack tip for $\bar a = 0.3$. As expected the stress diverges as $\bar x_1^{-\frac{1}{2}}$  close to the crack tip and the prediction matches well with the numerical solution for $10^{-4}<\bar x_1< 10^{-2}$. Far away from the crack tip, the perturbed stresses vanish.
The angular distribution of stresses obtained from the asymptotic solution agrees with that from the numerical solution in Figure \ref{sigma_theta_short} at $|\bar r - \bar a| = 0.012$. The distribution is somewhat similar to that obtained for the isotropic cases \citep{lawn90}.
Figure \ref{short_t_sigma} compares the angular distribution of the stresses at various radial distances from the crack tip, both in the short and the long time limits. The magnitude of the stresses at a given location in the long time limit are lower than those in the short time limit. This decrease may be attributed to the flow of the solvent that relieves any pressure variation that develops in the short time limit. Compared to the short time limit, the angular variation of the stress in the long time limit show larger deviations from the isotropic case as also suggested by the values of $\delta_i$ in (\ref{eq:4103}).

The Griffith's criteria (\ref{eq:41634}) shows that the pressure required to open a crack increases with decreasing size of the crack. Since the maximum dimensionless capillary pressure is about \cite{mason95} $ 5.3 $, the largest allowable flaw which will not crack the sample in the short time limit is, $\tilde a_{\mathrm{max}} = 0.025 \sqrt{W}$ which suggests that packings containing particles of larger size and/or higher shear moduli can resist cracking more effectively. Recently, Tirumkudulu and Russel\cite{mahesh05} have derived the expression for the critical capillary stress to drive an infinite crack through a drying colloidal thin film bound to a substrate, $(-\tilde P^o_{\infty}) (\tilde h^{2/3}) = 0.23 W^{1/3}$. Comparing the critical capillary pressure for the two cases suggests that when $\tilde a \ll \tilde h$, a significantly larger capillary pressure is required to expand a finite flaw in the film compared to that required to drive an infinite crack,
\[
\frac{\tilde P^o}{\tilde P^o_{\infty}}\sim \left( \frac{\tilde h }{\tilde a}\right )^{\frac{2}{3}}.
\] 

These results are in line with the recent theoretical results obtained by Russel et al.\cite{russel08a} using the more accurate constitutive relation and experiments measuring the critical capillary pressure for various particle packings\citep{russel08b}.

The energy release rate, $\mathcal{G} = 2J$, is related to the stress intensity factor through the standard relation, $\mathcal{G} = \frac{K^2}{E_{\mathrm{eff}}}$ where the effective elastic modulus for the packing,
\be
E_{\mathrm{eff}} = \frac{E}{Q^{-1}_{11}}\;,
\ee 
accounts for the particle size and packing, and also for the anisotropy resulting from the nucleation of crack.

It is important to note the limitations of the analysis presented here. The boundary condition $\sgp_{11} =  - \sigma^o_{11}$ at the surface of the crack implies perturbed strains of the order $\eo$ close to the crack surface, which is inconsistent with the linearization in (\ref{eq:4501}). The same applies to the diverging perturbed strains at the crack tip as predicted by the linear analysis. The extent of errors introduced by such approximations can be accurately determined only by solving numerically the full non-linear momentum balance equations. However, the recent experimental evidence of diverging stresses close to the tip of a crack in a colloidal packing supports the overall trend predicted by the linear analysis.

Finally, the analysis presented here is general, in that the results relating to the asymptotic forms of the stress and displacement component, and the related expression for the energy release rate can easily be obtained for any other constitutive equation for a saturated packed bed once the stiffness matrix for the linearized equation is known.
\section{Conclusions}
We present the asymptotic analysis of the deformation field near a crack tip for a mode I crack in a two dimensional colloidal packing saturated with solvent. The stress and strain fields are linearized about the pre-crack state to yield the stress intensity factor for the two dimensional elastic field which is then related to the surface energy using the well known Griffith's criterion for equilibrium cracks. The calculated quantities are then compared with the numerical solution for the full problem. The main findings can be summarized as follows:
\begin{itemize}
\item Perturbation in the displacement and stress field due to the presence of crack introduces anisotropy in the material which can be quantified by (\ref{eq:4103}).

\item The stress and displacement fields close to crack tip are given by (\ref{eq:41301}) where the expression of $\mathbf{B}$ is obtained from the components of stiffness matrix, $\mathbf{C}$.

\item The critical pressure required to open a flaw of length $2a$ varies inversely with the crack length to the two thirds' power,
\[ -P^o = A \left( {GM\rcp}\right)^{1/3} \left(\frac{2\gamma}{a}\right)^{2/3},\]
where $A$ is equal to 0.45 and 0.35 for the short and long time limit, respectively. It is independent of the particle size.
\item The maximum flaw size that can resist cracking and result in a crack-free packing is set by the maximum possible capillary pressure, 
\[a_{\mathrm{max}} = \left( \frac{A}{5.3} \right)^{3/2}   \left( \frac{GM\rcp R^3}{2 \gamma}\right)^{1/2}.  \] 
Colloidal beds containing large  particles with high shear modulus are less susceptible to cracking.
\item When $a \ll h$, the critical capillary pressure required to expand a flaw is much larger that that required to drive an infinite crack in a film of thickness, $h$,
\[
\frac{P^o}{ P^o_{\infty}}\sim \left( \frac{ h }{ a}\right )^{\frac{2}{3}}.
\] 

\end{itemize}

\begin{acknowledgments}
The research was financially supported in part by the Department of Science and Technology, India (Project \#07DS032). A.\ S.\ acknowledges IIT Bombay's support for teaching assistantship.
\end{acknowledgments}



%

\end{document}